# Development of magnetism in Fe-doped magnetic semiconductors: Resonant photoemission and x-ray magnetic circular dichroism studies of (Ga,Fe)As

Takahito Takeda[1], Shoya Sakamoto[2], Le Duc Anh[1,3,4], Yukiharu Takeda[5],
Shin-ichi Fujimori[5], Miho Kitamura[6], Koji Horiba[6], Hiroshi Kumigashira[6,7],
Atsushi Fujimori[8,9], Masaaki Tanaka[1,10], and Masaki Kobayashi[1,10*]

[1]*Department of Electrical Engineering and Information Systems, The University of Tokyo, 7-3-1 Hongo, Bunkyo-ku, Tokyo 113-8656, Japan*
[2]*The Institute for Solid State Physics, The University of Tokyo, 5-1-5 Kashiwanoha, Kashiwa, Chiba 277-8581, Japan*
[3]*Institute of Engineering Innovation. The University of Tokyo, 7-3-1 Hongo, Bunkyo-ku, Tokyo 113-0032, Japan*
[4]*PRESTO, Japan Science and Technology Agency, 4-1-8 Honcho, Kawaguchi, Saitama 332-0012, Japan*
[5]*Materials Sciences Research Center, Japan Atomic Energy Agency, Sayo-gun, Hyogo 679-5148, Japan*
[6]*Photon Factory, Institute of Materials Structure Science, High Energy Accelerator Research Organization (KEK), 1-1 Oho, Tsukuba 305-0801, Japan*
[7]*Institute of Multidisciplinary Research for Advanced Materials (IMRAM), Tohoku University, Sendai 980–8577, Japan*
[8]*Department of Physics, The University of Tokyo, 7-3-1 Hongo, Bunkyo-ku, Tokyo 113-0033, Japan*
[9]*Department of Applied Physics, Waseda University, Okubo, Shinjuku, Tokyo 169-8555, Japan*
[10]*Center for Spintronics Research Network, The University of Tokyo, 7-3-1 Hongo, Bunkyo-ku, Tokyo 113-8656, Japan*

*Author to whom all correspondence should be addressed: masaki.kobayashi@ee.t.u-tokyo.ac.jp

## ABSTRACT

Fe-doped III-V ferromagnetic semiconductors (FMSs) such as (In,Fe)As, (Ga,Fe)Sb, (In,Fe)Sb, and (Al,Fe)Sb are promising materials for spintronic device applications because of the availability of both n- and p-type materials and the high Curie temperatures. On the other hand, (Ga,Fe)As, which has the same zinc-blende crystal structure as the Fe-doped III-V FMSs, shows paramagnetism. The origin of the different magnetic properties in the Fe-doped III-V semiconductors remains to be elucidated. To address this issue, we use resonant photoemission spectroscopy (RPES) and x-ray magnetic circular dichroism (XMCD) to investigate the electronic and magnetic properties of the Fe ions in a paramagnetic $(Ga_{0.95},Fe_{0.05})As$ thin film. The observed Fe $2p$-$3d$ RPES spectra show that the Fe $3d$ states are similar to those of ferromagnetic (Ga,Fe)Sb. The estimated Fermi level is located in the middle of the band gap in (Ga,Fe)As. The Fe $L_{2,3}$ XMCD spectra of $(Ga_{0.95},Fe_{0.05})As$ show pre-edge structures, which are not observed in the Fe-doped FMSs, indicating that the minority-spin ($\downarrow$) $e_\downarrow$ states are vacant in $(Ga_{0.95},Fe_{0.05})As$. The XMCD results suggest that the carrier-induced ferromagnetic interaction in $(Ga_{0.95},Fe_{0.05})As$ is short-ranged and weaker than that in the Fe-doped FMSs. The experimental findings suggest that the electron occupancy of the $e_\downarrow$ states contributes to the appearance of ferromagnetism in the Fe-doped III-V semiconductors, for p-type as well as n-type compounds.

## I. INTRODUCTION

Ferromagnetic semiconductors (FMSs) are synthesized by doping magnetic ions in semiconductor hosts. In III-V FMSs, a sizable amount of magnetic ions, such as Mn and Fe ions, partially replaces the cation sites (group III sites). The magnetic interaction between the doped magnetic ions is considered to be mediated by the spin of the carriers, and such ferromagnetism is called carrier-induced ferromagnetism[1]. To explain the ferromagnetism of FMSs, two models have been proposed so far: Zener's *p-d* exchange model and the impurity band (IB) model. Itinerant carriers (band conduction)[2,3] mediate the ferromagnetic (FM) *p-d* exchange interaction in Zener's *p-d* exchange model, while localized carriers (hopping conduction)[4,5] mediate the FM double-exchange interaction in the IB model. Here, the position of the Fermi level ($E_F$) relative to the 3*d* IB and the valence band (VB) differs between these two models.

The Mn-doped III-V FMSs, such as (In,Mn)As[6,7,8,9] and (Ga,Mn)As[10,11,12], have been intensively studied for more than two decades. Nevertheless, they have the following problems to be solved for applications: The Curie temperature ($T_C$) of the Mn-doped FMSs is much lower than room temperature[13,14] and their carriers are only p-type. Recently, the Fe-doped III-V FMSs such as n-type (In,Fe)As[15,16,17], n-type (In,Fe)Sb[18,19,20], p-type (Ga,Fe)Sb[21,22,23], and insulating (Al,Fe)Sb[24] have been grown by molecular beam epitaxy (MBE). Since the valence of Fe ions at the cation sites of the III-V semiconductors is expected to be 3+ and conducting carriers can be additionally introduced, one can independently control the Fe and carrier concentrations in the Fe-doped FMSs. Furthermore, the highest $T_C$ values reported so far in $(In_{0.65},Fe_{0.35})Sb$ (385

K)[20] and $(Ga_{0.8},Fe_{0.2})Sb$ (> 400 K)[25] are well above room temperature. These advantages make the Fe-doped FMSs promising for practical devices operating at room temperature. The origin of the high $T_C$ in the Fe-dope FMSs has been recently studied. As for (Ga,Fe)Sb, some experimental[26,27,28,29] and theoretical[30,31] studies on its electronic structure have been performed to understand the mechanism of the high-$T_C$ carrier-induced ferromagnetism. From a series of the experimental studies, the following spectroscopic methods are found very useful to elucidate the origin of the ferromagnetism in (Ga,Fe)Sb: X-ray absorption spectroscopy (XAS), x-ray magnetic circular dichroism (XMCD), resonant photoemission spectroscopy (RPES)[26,29], and angle-resolved photoemission spectroscopy (ARPES)[27]. These studies suggest that the electronic structure of (Ga,Fe)Sb is consistent with the IB model and that both the double-exchange interaction and the $p$-$d$ exchange interaction contribute to the ferromagnetism. The electronic structures and magnetic properties of the other Fe-doped III-V FMSs, (Al,Fe)Sb[32] and (In,Fe)As[33,34,35], have been studied using the x-ray spectroscopic methods, and these studies have demonstrated the importance of clarifying the relationship between the electronic structure and the ferromagnetism in the Fe-doped III-V FMSs.

In order to fundamentally understand the magnetism in the Fe-doped III-V semiconductors, it is helpful to study a similar compound that does not show ferromagnetism. (Ga,Fe)As is paramagnetic (PM) when the Fe distribution is uniform, thus it is an ideal case of a paramagnetic Fe-doped III-V semiconductor. Note that (Ga,Fe)As with inhomogeneous Fe distribution shows FM features[36,37,38]. Theoretical studies[39,40] also suggest that (Ga,Fe)As does not show ferromagnetism but paramagnetism. The absence of ferromagnetism has been attributed to the weakness of the *p-d* hybridization[41,42]. The comparison of the electronic structure and magnetic properties between the PM and FM Fe-doped III-V semiconductors will provide a key to understand the origin of the ferromagnetism in the Fe-doped FMSs. In this paper, we investigate the electronic structure of (Ga,Fe)As using RPES and XMCD to reveal the relationship between the magnetism and the band structure in the Fe-doped III-V semiconductors. Our findings suggest that the ferromagnetism in the Fe-doped III-V FMSs possibly appears when the $e_\downarrow$ states are occupied by electrons, and $T_C$ rises with the increase of the carrier concentration. Note that, in this paper, "occupation" means that the states are occupied by electrons.

## II. EXPERIMENTAL

A $(Ga_{0.95},Fe_{0.05})As$ thin film with a thickness of 30 nm was grown on a semi-insulating (SI) GaAs(001) substrate by MBE. The surface of the film was covered by a thin GaAs capping layer (~2 nm) to avoid surface contamination. The sample structure is, from top to bottom, GaAs capping layer ~2 nm/$(Ga_{0.95},Fe_{0.05})As$ 30 nm/GaAs 100 nm/SI-GaAs substrate. The excellent crystallinity of the sample was confirmed by reflection high-energy electron diffraction (RHEED) during the MBE growth, as shown in Fig. 1(a), which is similar to that of previous studies[36,37].

The XAS and XMCD experiments were performed at beamline BL23SU of SPring-8. The measurements were conducted under an ultrahigh vacuum below $1.7 \times 10^{-8}$ Pa at a temperature of 10 K. The strength of applied magnetic field ($\mu_0 H$) was varied from -7 T to 7 T. The direction of the magnetic field was parallel to the incident x-ray and perpendicular to the sample surface. XAS and XMCD spectra were measured in the total-electron-yield (TEY) mode. XMCD was measured by reversing the helicity of x rays with 1-Hz frequency at each photon energy under a fixed magnetic field, and the scans were repeated with the opposite magnetic field direction. XMCD spectra were averaged as $\left(\left(\sigma_{+,h} - \sigma_{-,h}\right) - \left(\sigma_{+,-h} - \sigma_{-,-h}\right)\right)/2$ , and each XAS spectrum was obtained as $\left(\sigma_{+,h} + \sigma_{-,h}\right)/2$, where $\sigma_+$ and $\sigma_-$ represent circularly polarized x rays with the photon helicity parallel and antiparallel to the spin polarization, and $h$ and $-h$ mean the magnetic field directions.

The RPES measurements were performed at beamline BL2A of Photon Factory. The measurements were conducted under an ultrahigh vacuum below $5.0\times10^{-9}$ Pa at a temperature of 20 K. The photon energy ($h\nu$) of the incident beam was varied from 690 eV to 750 eV and horizontal polarization was used for the measurements. The total energy resolutions including the thermal broadening were between 180 meV and 230 meV depending on $h\nu$. X-ray photoemission spectroscopy (XPS) measurements were also performed with an Mg-$K\alpha$ light source ($h\nu = 1253.6$ eV) at room temperature, where the energy resolution was about 800 meV. The $E_F$ position has been corrected using the Fermi edge of Au in electrical contact with the samples and the binding energy ($E_B$) is measured relative to $E_F$.

For the XAS, XMCD, and XPS measurements, the samples were etched in hydrochloric acid (HCl) and rinsed in water just before loading the samples into the vacuum chambers[26,43] in order to obtain clean surfaces. As for the RPES experiments, we measured the unetched sample.

### III. RESULTS AND DISCUSSION

**A. Macroscopic magnetic properties of $(Ga_{0.95},Fe_{0.05})As$**

Reflection magnetic circular dichroism (MCD) spectroscopy has been conducted to investigate the magneto-optical properties of $(Ga_{0.95},Fe_{0.05})As$. Generally, the MCD intensity is proportional to the spin-splitting energy (Zeeman energy) of the host energy bands. Figure 1(b) shows normalized MCD spectra of the $(Ga_{0.95},Fe_{0.05})As$ thin film measured under various magnetic fields at $hv = 1 - 5$ eV. All these spectral line shapes are nearly identical and show a peak at the $E_1$ absorption edge ($hv = 2.73$ eV) of the host semiconductor GaAs. This result indicates that the MCD spectra originate from a single component in magnetism associated with the $sp$ bands of $(Ga_{0.95},Fe_{0.05})As$. Figure 1(c) shows the MCD – $H$ curves at various temperatures measured at the $E_1$ edge ($hv = 2.73$ eV). The $\mu_0 H$ dependence of the MCD intensity is linear, indicating that $(Ga_{0.95},Fe_{0.05})As$ is PM.

The carrier concentration of the $(Ga_{0.95},Fe_{0.05})As$ thin film cannot be estimated by a Hall measurement because the resistivity was too high. This suggests that the carrier concentration of $(Ga_{0.95},Fe_{0.05})As$ is much smaller than that of the Fe-doped FMSs including insulating $(Al,Fe)Sb$[24].

## B. XAS and XMCD spectra at the Fe $L_{2,3}$ absorption edge

Figure 2(a) shows XAS spectra at the Fe $L_{2,3}$ absorption edge of $(Ga_{0.95},Fe_{0.05})As$ at room temperature without magnetic field before etching (green curve), and under $T = 10$ K and $\mu_0 H = 7$ T after etching surface oxides by HCl (purple curve), where black dashed curves represent the background component. The XAS spectrum before etching has a peak at $\sim$710 eV besides the main peak at $\sim$708 eV. According to previous studies[26,29,43], the peak component at 710 eV is likely to originate from extrinsic $Fe^{3+}$ oxides. The comparison between the XAS spectra before and after etching suggests that the extrinsic $Fe^{3+}$ oxide component near the surface is almost removed by the etching process.

Figures 2(b) and 2(c) show the XAS and XMCD spectra at the Fe $L_{2,3}$ absorption edges of $(Ga_{0.95},Fe_{0.05})As$ measured at $T$ = 10 K under $\mu_0 H$ = 1, 4, and 7 T, where the intensities are normalized to the peak height at 708 eV in (c). The line shape of the XMCD spectra changes with increasing $\mu_0 H$, while the line shape of the XAS spectra is independent of $\mu_0 H$. It should be noted here that there is a pre-edge structure around ∼706 eV in the XMCD spectra taken at high magnetic fields indicated by the black arrow in Fig. 2(c), which is hardly seen in the XMCD spectra of the Fe-doped FMSs, $(In,Fe)As$[34,35], $(Al,Fe)Sb$[32], and $(Ga,Fe)Sb$[26]. This characteristic feature in $(Ga,Fe)As$ is discussed considering the VB structure below. As shown in Fig. 2(c), the ratio of the XMCD intensity at the $L_3$ edge to the $L_2$ edge changes significantly from $\mu_0 H$ = 1 T to $\mu_0 H$ = 4 T, indicating that the XMCD spectra are composed of multiple magnetic components. The change of the ratio is discussed below from the viewpoint of the spin and orbital magnetic moments.

In order to estimate the spin and orbital magnetic moments of the Fe ions in $(Ga,Fe)As$, we have applied the XMCD sum rules[44]:

$$m_{orb} = -\frac{2\int_{L_{2,3}}\text{XMCD}d\omega}{3\int_{L_{2,3}}\text{XAS}d\omega}n_h,$$

$$m_{spin} = -\frac{3\int_{L_3}\text{XMCD}d\omega - 2\int_{L_{2,3}}\text{XMCD}d\omega}{\int_{L_{2,3}}\text{XAS}d\omega}n_h,$$

where $m_{orb}$ and $m_{spin}$ are the orbital and spin magnetic moments in units of $\mu_B$, respectively, and $n_h$ is the number of $3d$ holes. Here, assuming that the valence of Fe is 3+, $n_h$ was set to 5. The correction factor of 0.685[45] for the $Fe^{3+}$ ion was used to estimate $m_{spin}$. The values of $m_{spin}$ and $m_{orb}$ under $\mu_0 H = 1$ T, 4 T, and 7 T at $T = 10$ K estimated from the XMCD sum rules are plotted as a function of $\mu_0 H$ in Fig. 3(a). As described below, the results of the sum rules are used to convert the XMCD – $H$ curves to the magnetization curves. Figure 3(b) shows the $\mu_0 H$ dependence of $m_{spin}/m_{orb}$. Figure 3(a) indicates that the intensity of XMCD increases with increasing $\mu_0 H$. The linear-$H$ component in Fig. 3(a), particularly at large $\mu_0 H$, would be the PM component because paramagnetism keeps increasing with $\mu_0 H$, while superparamagnetism and/or ferromagnetism eventually saturate at high magnetic fields. In contrast, in the other Fe-doped FMSs, the spectral line shapes of the XMCD spectra are unchanged with $\mu_0 H$, indicating that $m_{spin}/m_{orb}$ is independent of $\mu_0 H$.

## C. XMCD-*H* curves

Figure 4(a) shows the XAS and XMCD spectra, where we have indicated photon energies at which the $\mu_0 H$ dependence of the XMCD intensity (XMCD – *H* curve) was measured. Figure 4(b) shows XMCD – *H* curves at various photon energies, which have been normalized such that the slope of the curve between 0 T and 0.4 T coincides with each other in order to compare the line shape of the XMCD – *H* curves with that of the MCD – *H* curve. The XMCD – *H* curve taken at $hv$ = 707.7 eV, i.e., at the $L_3$ peak, is substantially different from the other curves as expected from the magnetic-field dependence of the XMCD spectra. The inset of Fig. 4(b) shows that the normalized XMCD – *H* curves change their slope at $\mu_0 H$ = ~0.5 T, unlike the linear $\mu_0 H$ dependence of the MCD at $E_1$ up to $\mu_0 H$ ~ 1 T, as shown in Fig. 1(c). Since the MCD reflects the Zeeman splitting of the *sp* bands and the XMCD reflects the spin polarization of Fe 3*d* electrons, the different features of the MCD – *H* and XMCD – *H* curves suggest that the Zeeman splitting is not proportional to the spin polarization of Fe 3*d* electrons. In contrast, in the case of the Fe-doped FMSs with a certain amount of carriers, the XMCD – *H* curves are nearly same as the MCD – *H* curves[26,35]. Thus, these results indicate that magnetic coupling between Fe $3d$ and As $4p$ electrons is not strong in the paramagnetic

(Ga$_{0.95}$,Fe$_{0.05}$)As.

## D. Fitting of the magnetization curve

As discussed above, the $\mu_0 H$ dependence of the XMCD suggests that multiple

magnetic components are present in (Ga$_{0.95}$,Fe$_{0.05}$)As. First-principles calculation study[46]

demonstrates that spin-glass states are stable for Fe ions in (Ga,Fe)As if the Ga sites are

randomly replaced by Fe ions. However, spatially inhomogeneous distribution of the Fe

ions induces FM Fe-Fe coupling: An interstitial Fe ion and the nearest-neighbor

substitutional Fe ion are ferromagnetically coupled[47]. Theoretical calculations for

(In,Fe)Sb and (Ga,Fe)Sb suggest that the cation-site Fe ion and the second-nearest-neighbor Fe ion make FM coupling[30]. Furthermore, non-uniform distribution of the Fe ions called spinodal decomposition tends to occur in Fe-doped FMSs[48]. Considering the line shapes of the XMCD – $H$ curves, we assume that the magnetism of (Ga$_{0.95}$,Fe$_{0.05}$)As can be decomposed into the PM and superparamagnetic (SPM) components, the latter of which is expected to originate from Fe-rich FM domains on the nm scale.

To characterize the FM domains in detail, the magnetization curve obtained from the XMCD-$H$ taken at $hv = 707.7$ eV is fitted by the linear combination[43,32,34] of the Langevin function $L(\xi)$ representing SPM behavior and a linear function representing PM one:

$$M = x m_{sat} L\left(\frac{\mu \mu_0 H}{k_{\mathrm{B}} T}\right) + (1 - x)\frac{C \mu_0 H}{T + T_A},$$

$$C = \frac{m_{sat}(m_{sat} + 2\mu_{\mathrm{B}})}{3 k_{\mathrm{B}}},$$

$$L(\xi) = coth(\xi) - \frac{1}{\xi},$$

where $M$ is the magnetization per Fe atom, $m_{sat}$ is the total magnetic moment of the Fe atom, $C$ is the Curie constant, and $T_A$ is the Weiss temperature. Here, $m_{sat}$ is set to 5 $\mu_{\mathrm{B}}$, the $g$ factor is 2 for simplicity, and $T_A$ is 32 K[36]. Fitting parameters are the following: $\mu$ is the total magnetic moment of a nanoscale FM domain, and $x$ is the fraction of Fe atoms in the domain. As shown in Fig. 4(c), the fitting well reproduces the experimental curve at $hv = 707.7$ eV using the $x$ and $\mu$ values of $0.10 \pm 0.01$ and $42 \pm 5$ $\mu_{\mathrm{B}}$, respectively. Note that the value of $x$ is roughly estimated or has qualitatively less physical meaning in this analysis since the $\mu_0 H$ dependence of the XMCD intensity depends on $h\nu$ reflecting the change of the XMCD spectral line shape with $\mu_0 H$ as shown in the inset of Fig. 2(c). That is, $x$ depends on $h\nu$ reflecting the different magnetic behavior between the SPM and PM components. On the other hand, the estimation of $\mu$ is approximately independent of $h\nu$ since $\mu$ is contained only in the Langevin function and the slopes of the XMCD – $H$ curves with low magnetic fields are nearly identical as shown in the inset of Fig. 4 (b). The magnetic moment per FM domain $\mu$ is roughly $40\mu_B$, which corresponds to 8 Fe atoms on average. This means that the average number of the Fe ions in the domain in $(Ga_{0.95}, Fe_{0.05})As$ is an order of magnitude smaller than that in the Fe-doped FMSs $((Ga, Fe)Sb, (Al, Fe)Sb)$ with nearly 5% of the Fe concentration[34,32]. The less Fe ions in the domain[49] indicates that the SPM component predominantly originates from the short-range magnetic interaction among the Fe ions in $(Ga_{0.95}, Fe_{0.05})As$, as compared with the long-range magnetic interaction in the Fe-doped FMSs.

**E. RPES spectra at the Fe $L_3$ absorption edge**

Figure 5(a) shows the $h\nu$ dependence of the RPES spectra in the unetched (Ga,Fe)As thin film, where the colors of the spectra correspond to the markers on the XAS spectrum

in the right panel. Here, the black spectrum in Fig. 5(a) is the off-resonant spectrum. To

extract the resonant behavior, the second-derivative image of the RPES spectra, which

are obtained by subtracting the off-resonant spectrum from the spectra in Fig. 5(a), is

shown in Fig. 5(b). Since the resonant enhancement in the vicinity of $E_F$ has not been

observed within the experimental accuracy, the Fe 3$d$ states are not located near $E_F$. It

should be noted here that no Auger components are seen in the RPES image of

$(Ga_{0.95},Fe_{0.05})As$. On the other hand, in the previous RPES study on $(Ga,Fe)Sb$[26], the

RPES image of $(Ga,Fe)Sb$ contains an Auger component below $E_B = 4$ eV. Generally,

Auger components in RPES measurements are clearly observed in metallic materials[50,51].

Thus, this comparison indicates that the Fe 3$d$ electrons in $(Ga_{0.95},Fe_{0.05})As$ are more localized than that in $(Ga,Fe)Sb$ because the intensity of the Auger components reflects the resonantly-enhanced density of states near $E_F$.

It should be noted here that the resonant behavior observed by RPES involves both

the Fe component in $(Ga,Fe)As$ and the oxidized Fe component, because the $(Ga,Fe)As$

film used in this RPES measurement was before HCl etching. To elucidate the origins of the resonant enhancement, it is useful to plot constant-initial-state (CIS) spectra which

are the intensity plot at fixed binding energies as functions of $h\nu$. Figure 5(c) shows the

CIS spectra at $E_B$ = 1.0 eV, 3.6 eV, and 6.6 eV, and the XAS spectra before and after

HCl etching. Since the CIS spectrum at $E_B$ = 6.6 eV (yellow curve in Fig. 5(c)) has a peak around $h\nu$ = 710 eV that can be removed by the HCl etching as described above,

the RPES spectra taken around $h\nu$ = 710 eV predominantly originate from the extrinsic

component. Here, since the peak around $h\nu$ = 708.4 eV of the CIS spectrum at $E_B$ = 6.6

eV comes from the large tail of the enhancement at $E_B$ = 3.6 eV (green curve in Fig.

5(c)), the peak around $h\nu$ = 708.4 eV is not attributed to the extrinsic component. The

position of the main peak of the after-etching XAS spectrum ($h\nu$ = 707.7 eV in the violet

curve in Fig. 5(c)) is close to the peak position of the CIS spectrum at $E_B$ = 1.0 eV ($h\nu$

= 707.9 eV, red-dot curve in Fig. 5(c)) rather than that at $E_B$ = 3.6 eV ($h\nu$ = 708.4 eV,

green-dot curve in Fig. 5(c)), indicating that the RPES spectrum taken at $h\nu$ = 707.7 eV

(red thick curve in Fig. 5(a)) predominantly reflects the partial density-of-states (PDOS)

of the substitutional Fe ions in (Ga,Fe)As. Based on the similarity of the Fe $3d$ electronic structure between (Ga,Fe)As and (Ga,Fe)Sb[26] as shown in the RPES mapping in Fig.

5(b), the resonantly-enhanced state at $E_B = 1.0$ eV (see the red dash line in Fig.5 (a)) is assigned as the $t_{2b\downarrow}$ states[29] as shown in Fig. 6 (b), where $t_{2b}$ and $\downarrow$ means the bonding state formed by the $p$-$d(t_2)$ hybridization and minority-spin, respectively.

To precisely estimate the position of the $t_{2b\downarrow}$ states, the Fe $3d$ PDOS spectrum (grey curve) taken at $h\nu = 707.9$ eV, which are obtained by subtracting the off-resonant spectrum (black curve in Fig. 5(a)) from the on-resonance spectrum taken at $h\nu = 707.9$ eV (thick red curve in Fig. 5(a)), is fitted by a Gaussian (red curve) and an asymmetric Gaussian (green curve) functions[26,29] as shown in Fig. 5(d). Here, the asymmetric component likely comes from the bonding state between the Fe $3d$ states and the ligand As $4p$ bands. The fitting result well reproduces the PDOS spectrum. No Fermi cutoff is observed, as shown in Fig. 5(d). On the other hand, in the previous study on (Ga,Fe)Sb[29], a clear Fermi cutoff is observed.

**F. Estimation of the valence band maximum by XPS**

To estimate the position of the valence band maximum (VBM), we have conducted XPS measurements with the Mg-$K\alpha$ line ($h\nu = 1253.6$ eV) on the (Ga$_{0.95}$,Fe$_{0.05}$)As film and a $(Ga_{0.95},Fe_{0.05})$Sb thin film, where the $(Ga_{0.95},Fe_{0.05})$Sb film is the same sample as that used in the previous study[29]. Figure 7 shows the XPS spectra of the VB. The values of the VBM of $(Ga_{0.95},Fe_{0.05})$As and $(Ga_{0.95},Fe_{0.05})$Sb are estimated to be 0.65 eV and 0.29 eV below $E_F$, respectively. Assuming that the band gap ($E_g$) of $(Ga_{0.95},Fe_{0.05})$As is the same as that of its host semiconductor GaAs ($E_g = 1.42$ eV[52]), $E_F$ is located in the middle of the band gap in $(Ga_{0.95},Fe_{0.05})$As. The result is consistent with the insulating nature of $(Ga_{0.95},Fe_{0.05})$As confirmed by the transport measurements.

Since the peak positions of the $t_{2b\downarrow}$ states of $(Ga_{0.95},Fe_{0.05})$As and $(Ga_{0.95},Fe_{0.05})$Sb are $E_B = 1.0$ eV (see Fig. 5(d)) and 1.6 eV[29], the energy differences between VBM and the $t_{2b\downarrow}$ states of $(Ga_{0.95},Fe_{0.05})$As and $(Ga_{0.95},Fe_{0.05})$Sb are 0.35 eV and 1.31 eV, respectively (see Fig. 6). Considering that the energy difference between the VBM and bonding states is proportional to the strength of the hybridization, this smaller value of the energy difference in $(Ga_{0.95},Fe_{0.05})$As than that in $(Ga_{0.95},Fe_{0.05})$Sb suggests that the *p-d* hybridization in (Ga,Fe)As is weaker than that in (Ga,Fe)Sb. As mentioned above (section C), the disagreement between the XMCD – *H* and MCD – *H* curves also suggests the weakness of the *p-d* hybridization in (Ga,Fe)As.

**G. Discussion**

Based on the experimental findings, let us discuss the essence of the magnetism in the Fe-doped III-V semiconductors. The Fe $L_{2,3}$ XMCD spectra of the Fe-doped FMSs do not show the pre-edge structure observed in $(Ga_{0.95},Fe_{0.05})As$, as shown in Fig. 2(c)[26,32]. In contrast to the Fe-doped FMSs, the pre-edge structure of the XMCD spectra has been observed in (Ga,Mn)As[53,54]. In general, since X-ray absorption occurs from core-level electrons to the unoccupied states, the pre-edge structures of XAS and XMCD reflect the unoccupied states in the vicinity of $E_F$. For instance, as for (Ga,Mn)As, the pre-edge structure is considered to originate from the unoccupied $p$-$d$ hybridized state[53]. The pre-edge structure in the XMCD spectrum of $(Ga_{0.95},Fe_{0.05})As$ can be seen at high magnetic fields ($\mu_0 H > 1$ T), while the intensity of the pre-edge structure is probably too weak to be observed at $\mu_0 H = 1$ T. Here, the XMCD spectra under $\mu_0 H = 1$ T mainly reflect the SPM feature which is expected from Fe-rich FM domains in nm scale. Thus, the pre-edge structure comes from the PM component. Since the Fe $3d$ electronic structure of $(Ga_{0.95},Fe_{0.05})As$ is similar to that of (Ga,Fe)Sb, the Fe $3d$ states related to the pre-edge structure would be elucidated by comparing $(Ga_{0.95},Fe_{0.05})As$ with $(Ga,Fe)Sb$. In the case of $(Ga,Fe)Sb$, the $e_\downarrow$ states are partially occupied and the $t_{2a\downarrow}$ states are vacant[29] as shown in Fig. 6(a), where subscript $a$ means antibonding. As for $(Ga,Fe)Sb$, the $e_\downarrow$ states are gradually occupied with the increase of the Fe concentration, since the overlap between the $e_\downarrow$ states and the $t_{2a\uparrow}$ states increases due to the broadening of the Fe $3d$ states[27,29] (see Fig. 6(a)). When the Fe concentration increases, $(Ga,Fe)As$ would show the same trend of the $e_\downarrow$-state occupancy as $(Ga,Fe)Sb$. The $\mu_0 H$ dependence of the pre-edge structure of $(Ga_{0.95},Fe_{0.05})As$ can be explained by the difference of the occupancy of the $e_\downarrow$ states between the SPM and PM components. It is likely that the absence of the pre-edge structure in $(Ga,Fe)Sb$ results from the partial occupation of the $e_\downarrow$ states, as shown in Fig. 6(a). The $e_\downarrow$ states of the SPM component are expected to be slightly occupied in $(Ga_{0.95},Fe_{0.05})As$ like the partial occupation of the $e_\downarrow$ states in $(Ga,Fe)Sb$ as shown in Fig. 6(a). Taking into account that there is no resonant enhancement in the vicinity of $E_F$ in $(Ga_{0.95},Fe_{0.05})As$, as shown in Fig. 5(d), the $e_\downarrow$ states of $(Ga_{0.95},Fe_{0.05})As$ is probably vacant in the PM region and the pre-edge structure would reflect the unoccupied $e_\downarrow$ states. The Fe $3d$ electronic structure of $(Ga_{0.95},Fe_{0.05})As$ of the PM region is schematically shown in Fig. 6(b).

The comparison of the electronic structures between $(Ga_{0.95},Fe_{0.05})As$ and $(Ga,Fe)Sb$ indicates that the occupation of the $e_\downarrow$ states is the key element of ferromagnetism in Fe-doped FMSs. The difference of the $e_\downarrow$-state occupancy between PM $(Ga_{0.95},Fe_{0.05})As$ and FM $(Ga,Fe)Sb$ likely reflects the difference of the strength of the $p$-$d$ hybridization: Since the energy separation between the VBM and the $t_{2a\uparrow}$ states is proportional to the strength of the $p$-$d$ hybridization, the overlap between the $e_\downarrow$ states and $t_{2a\uparrow}$ states, which possibly leads to the electron occupation in the $e_\downarrow$ state (in the p-type cases), depends on the $p$-$d$ hybridization[27,29].

To decompose the role of the carriers and Fe $3d(e_\downarrow)$ electrons, let us apply the above-mentioned discussion on $(Ga_{0.95},Fe_{0.05})As$ to $(Al,Fe)Sb$ and $(Ga,Fe)Sb$ quantum wells. The previous study on $(Al,Fe)Sb$[24] demonstrates that $(Al,Fe)Sb$ is insulating at low temperatures although Hall measurements can be conducted at room temperature, and it is considered that the ferromagnetism mainly originates from the short-range interaction in the Fe-rich domains due to the low carrier concentration. The XMCD study on (Al,Fe)Sb suggests that Fe ions take the $3d^6$ configuration with a ligand hole[32]. From the first-principles calculation[31], (Al,Fe)Sb has partially-occupied $e_\downarrow$ states. Since the $T_C$ and carrier concentration of (Al,Fe)Sb are lower than those of (Ga,Fe)Sb with the same Fe concentration[22,24], $T_C$ probably rises with the increase of the $e_\downarrow$-state occupation under the condition of the constant Fe concentration. Moreover, the previous study on (Ga,Fe)Sb quantum wells[22] demonstrates that $T_C$ of the (Ga,Fe)Sb quantum wells falls with the decrease of the well thickness. The drop in the $T_C$ of the (Ga,Fe)Sb quantum wells can be explained by the decrease of the carrier concentration due to a carrier depletion layer at the interface between the AlSb buffer and the (Ga,Fe)Sb quantum-well layers[22]. The decrease of the carrier concentration leads to the decrease of the $e_\downarrow$-state occupation. Note that the behavior of the carrier concentration and that of the $e_\downarrow$-state occupation are different from those of the usual "p-type" materials depending on the strength of the *p-d* hybridization, since the carriers are derived from the (relatively) wide $t_{2a\uparrow}$ band rather than the localized $e_\downarrow$ states. Additionally, a previous SX-ARPES study on n-type (In,Fe)As has revealed that the partial occupation of the $e_\downarrow$ state results in the large *s-d* exchange interaction, leading to the electron-carrier induced ferromagnetism[33]. This means, in the n-type Fe-doped FMSs, the occupation of the $e_\downarrow$ states is indispensable for the development of ferromagnetism as in the case of the p-type ones. Therefore, it follows from these arguments that ferromagnetism develops in the Fe-doped FMSs, for both p-type and n-type, when the $e_\downarrow$ states are partially occupied.

## IV. SUMMARY

We have performed RPES and XMCD measurements on PM $(Ga_{0.95},Fe_{0.05})As$ thin films to unveil the origin of the magnetism in Fe-doped III-V semiconductors. The RPES spectra of $(Ga_{0.95},Fe_{0.05})As$ reveal that the Fe $3d$ states are similar to those of (Ga,Fe)Sb except for the Auger component. The absence of the Auger component indicates that the Fe $3d$ electrons are localized in (Ga,Fe)As. The estimated Fermi level is located in the middle of the band gap in the $(Ga_{0.95},Fe_{0.05})As$ film, consistent with the fact that the carrier concentration is too low to perform Hall measurements. The Fe XMCD spectra of $(Ga_{0.95},Fe_{0.05})As$ show peaks in the pre-edge region, which are not observed in the Fe-doped FMSs. We argue that this pre-edge features originate from the localized unoccupied $e$ states just above $E_F$. The XMCD results suggest that the short-range magnetic interaction among the Fe ions is dominant in $(Ga_{0.95},Fe_{0.05})As$ unlike in the Fe-doped FMSs. The small energy difference between the VBM and the $t_{2b\downarrow}$ states of $(Ga_{0.95},Fe_{0.05})As$ suggests the $p$-$d$ hybridization in (Ga,Fe)As is weaker than that in (Ga,Fe)Sb. The unoccupied $e_\downarrow$ states originate from the weak $p$-$d$ hybridization. Based on the experimental findings, we conclude that the occupancy of the $e_\downarrow$ states contributes to the appearance of the ferromagnetism in the Fe-doped III-V semiconductors, not only for n-type but also for p-type compounds, and that $T_C$ rises with the increase of the $e_\downarrow$-state occupation. It has been believed that the itinerant Fe $3d$ electrons mainly contribute to the ferromagnetism in the FMSs so far. This study suggests that, in addition to the $t_2$ states that well hybridize with the ligand, the relatively localized $e$ states also play an important role for the ferromagnetism in the Fe-doped FMSs.

**ACKNOWLEWDGMENTS**

This work was supported by Grants-in-Aid for Scientific Research (No. 15H02109, No. 16H02095, No. 19K21961, No. 18H05345, No. 20H05650 and No. 23000010). This work was supported by JST-CREST (JPMJCR18T5 and JPMJCR1777). L. D. A.

acknowledge the support from PRESTO Program (JPMJPR19LB) of Japan Science and Technology Agency. This work was partially supported by the Spintronics Research Network of Japan (Spin-RNJ). This work was performed under the Shared Use Program of Japan Atomic Energy Agency (JAEA) Facilities (Proposal No. 2017B-E19 and 2019A-E15) supported by JAEA Advanced Characterization Nanotechnology Platform as a program of "Nanotechnology Platform" of the Ministry of Education, Culture, Sports, Science and Technology (MEXT) (Proposal No. A-17-AE-0038 and A-19-AE-0015). Supporting experiments at SPring-8 were approved by the Japan Synchrotron Radiation Research Institute (JASRI) Proposal Review Committee (Proposal No. 2017B3841 and No. 2019A3841). This work at KEK-PF was performed under the approval of the Program Advisory Committee (Proposals 2018G114 and 2020G112) at the Institute of Materials Structure Science at KEK.

# REFERENCES

[1] T. Dietl, A. Haury, and Y. Merle d'Aubigné, Phys. Rev. B **55**, R3347 (1997).

[2] T. Dietl, H. Ohno, F. Matsukura, J. Cibert, and D. Ferrand, Science (80-. ). **287**, 1019 (2000).

[3] T. Dietl, H. Ohno, and F. Matsukura, Phys. Rev. B **63**, 195205 (2001).

[4] M. Berciu and R.N. Bhatt, Phys. Rev. Lett. **87**, 107203 (2001).

[5] S. Ohya, I. Muneta, Y. Xin, K. Takata, and M. Tanaka, Phys. Rev. B **86**, 094418 (2012).

[6] H. Munekata, H. Ohno, S. von Molnar, A. Segmüller, L.L. Chang, and L. Esaki, Phys. Rev. Lett. **63**, 1849 (1989).

[7] H. Ohno, H. Munekata, S. von Molnár, and L.L. Chang, J. Appl. Phys. **69**, 6103 (1991).

[8] H. Munekata, H. Ohno, R.R. Ruf, R.J. Gambino, and L.L. Chang, J. Cryst. Growth **111**, 1011 (1991).

[9] H. Ohno, H. Munekata, T. Penney, S. von Molnár, and L.L. Chang, Phys. Rev. Lett. **68**, 2664 (1992).

[10] H. Ohno, A. Shen, F. Matsukura, A. Oiwa, A. Endo, S. Katsumoto, and Y. Iye, Appl. Phys. Lett. **69**, 363 (1996).

[11] A. Van Esch, L. Van Bockstal, J. De Boeck, G. Verbanck, A.S. van Steenbergen, P.J. Wellmann, B. Grietens, R. Bogaerts, F. Herlach, and G. Borghs, Phys. Rev. B **56**, 13103 (1997).

[12] T. Hayashi, M. Tanaka, T. Nishinaga, H. Shimada, H. Tsuchiya, and Y. Otuka, J. Cryst.

Growth **175**–**176**, 1063 (1997).

[13] L. Chen, X. Yang, F. Yang, J. Zhao, J. Misuraca, P. Xiong, and S. von Molnár, Nano Lett. **11**, 2584 (2011).

[14] T. Schallenberg and H. Munekata, Appl. Phys. Lett. **89**, 042507 (2006).

[15] P. Nam Hai, D. Sasaki, L.D. Anh, and M. Tanaka, Appl. Phys. Lett. **100**, 262409 (2012).

[16] P. Nam Hai, L. Duc Anh, S. Mohan, T. Tamegai, M. Kodzuka, T. Ohkubo, K. Hono, and M. Tanaka, Appl. Phys. Lett. **101**, 182403 (2012).

[17] P. Nam Hai, L.D. Anh, and M. Tanaka, Appl. Phys. Lett. **101**, 252410 (2012).

[18] A. V Kudrin, Y.A. Danilov, V.P. Lesnikov, M. V Dorokhin, O. V Vikhrova, D.A. Pavlov, Y. V Usov, I.N. Antonov, R.N. Kriukov, A. V Alaferdov, and N.A. Sobolev, J. Appl. Phys. **122**, 183901 (2017).

[19] N.T. Tu, P.N. Hai, L.D. Anh, and M. Tanaka, Appl. Phys. Express **11**, 063005 (2018).

[20] N.T. Tu, P.N. Hai, L.D. Anh, and M. Tanaka, Appl. Phys. Express **12**, 103004 (2019).

[21] N.T. Tu, P.N. Hai, L.D. Anh, and M. Tanaka, Appl. Phys. Lett. **105**, 132402 (2014).

[22] N.T. Tu, P.N. Hai, L.D. Anh, and M. Tanaka, Phys. Rev. B **92**, 144403 (2015).

[23] N.T. Tu, P.N. Hai, L.D. Anh, and M. Tanaka, Appl. Phys. Lett. **108**, 192401 (2016).

[24] L.D. Anh, D. Kaneko, P.N. Hai, and M. Tanaka, Appl. Phys. Lett. **107**, 232405 (2015).

[25] S. Goel, L.D. Anh, S. Ohya, and M. Tanaka, Phys. Rev. B **99**, 014431 (2019).

[26] S. Sakamoto, N.T. Tu, Y. Takeda, S. Fujimori, P.N. Hai, L.D. Anh, Y.K. Wakabayashi, G. Shibata, M. Horio, K. Ikeda, Y. Saitoh, H. Yamagami, M. Tanaka, and A. Fujimori, Phys. Rev. B **100**, 035204 (2019).

[27] T. Takeda, M. Suzuki, L.D. Anh, N.T. Tu, T. Schmitt, S. Yoshida, M. Sakano, K. Ishizaka, Y. Takeda, S. Fujimori, M. Seki, H. Tabata, A. Fujimori, V.N. Strocov, M. Tanaka, and M. Kobayashi, Phys. Rev. B **101**, 155142 (2020).

[28] K. Sriharsha, L.D. Anh, N.T. Tu, S. Goel, and M. Tanaka, APL Mater. **7**, 021105 (2019).

[29] T. Takeda, S. Sakamoto, K. Araki, Y. Fujisawa, L.D. Anh, N. Thanh Tu, Y. Takeda, S. Fujimori, A. Fujimori, M. Tanaka, and M. Kobayashi, Phys. Rev. B **102**, 245203 (2020).

[30] H. Shinya, T. Fukushima, A. Masago, K. Sato, and H. Katayama-Yoshida, J. Appl. Phys. **124**, 103902 (2018).

[31] S. Sakamoto and A. Fujimori, J. Appl. Phys. **126**, 173910 (2019).

[32] S. Sakamoto, L.D. Anh, P.N. Hai, Y. Takeda, M. Kobayashi, Y.K. Wakabayashi, Y. Nonaka, K. Ikeda, Z. Chi, Y. Wan, M. Suzuki, Y. Saitoh, H. Yamagami, M. Tanaka, and A. Fujimori, Phys. Rev. B **101**, 075204 (2020).

[33] M. Kobayashi, L.D. Anh, J. Minár, W. Khan, S. Borek, P.N. Hai, Y. Harada, T. Schmitt, M. Oshima, A. Fujimori, M. Tanaka, and V.N. Strocov, Phys. Rev. B **103**, 115111 (2021).

[34] S. Sakamoto, L.D. Anh, P.N. Hai, G. Shibata, Y. Takeda, M. Kobayashi, Y. Takahashi, T. Koide, M. Tanaka, and A. Fujimori, Phys. Rev. B **93**, 035203 (2016).

[35] M. Kobayashi, L.D. Anh, P.N. Hai, Y. Takeda, S. Sakamoto, T. Kadono, T. Okane, Y. Saitoh, H. Yamagami, Y. Harada, M. Oshima, M. Tanaka, and A. Fujimori, Appl. Phys. Lett. **105**, 032403 (2014).

[36] S. Haneda, M. Yamaura, Y. Takatani, K. Hara, S. Harigae, and H. Munekata, Jpn. J. Appl. Phys. **39**, L9 (2000).

[37] R. Moriya, Y. Katsumata, Y. Takatani, S. Haneda, T. Kondo, and H. Munekata, Phys. E Low-Dimensional Syst. Nanostructures **10**, 224 (2001).

[38] A. V Kudrin, V.P. Lesnikov, Y.A. Danilov, M. V Dorokhin, O. V Vikhrova, P.B. Demina, D.A. Pavlov, Y. V Usov, V.E. Milin, Y.M. Kuznetsov, R.N. Kriukov, A.A. Konakov, and N.Y. Tabachkova, Semicond. Sci. Technol. **35**, 125032 (2020).

[39] L.M. Sandratskii and P. Bruno, Phys. Rev. B **67**, 214402 (2003).

[40] H. Katayama-Yoshida and K. Sato, J. Phys. Chem. Solids **64**, 1447 (2003).

[41] S. Mirbt, B. Sanyal, C. Isheden, and B. Johansson, Phys. Rev. B **67**, 155421 (2003).

[42] W. Shu-Yi, W. Tian-Xing, Y. Zong-Xian, and M. Li, Commun. Theor. Phys. **40**, 499 (2003).

[43] Y.K. Wakabayashi, S. Sakamoto, Y. Takeda, K. Ishigami, Y. Takahashi, Y. Saitoh, H. Yamagami, A. Fujimori, M. Tanaka, and S. Ohya, Sci. Rep. **6**, 23295 (2016).

[44] B.T. Thole, P. Carra, F. Sette, and G. van der Laan, Phys. Rev. Lett. **68**, 1943 (1992).

[45] Y. Teramura, A. Tanaka, and T. Jo, J. Phys. Soc. Japan **65**, 1053 (1996).

[46] K. Sato and H. Katayama-Yoshida, Semicond. Sci. Technol. **17**, 367 (2002).

[47] L.D. Anh, T. Hayakawa, Y. Nakagawa, H. Shinya, T. Fukushima, M. Kobayashi, H. Katayama-Yoshida, Y. Iwasa, and M. Tanaka, Nat. Commun. **12**, 4201 (2021).

[48] T. Dietl, K. Sato, T. Fukushima, A. Bonanni, M. Jamet, A. Barski, S. Kuroda, M. Tanaka, P.N. Hai, and H. Katayama-Yoshida, Rev. Mod. Phys. **87**, 1311 (2015).

[49] The cubic unit cell of the zinc-blende crystal structure contains four group-III ions. If Fe ions replace one half of the group-III sites in the unit cell, 8 Fe ions occupy the volume of four-unit cells. Considering the average number of Fe ions estimated from the XMCD analysis, the minimal (spherical) domain size is ~1.1 nm. Here, the lattice constant of $(Ga_{0.95},Fe_{0.05})As$ is 5.656 Å [36]. Actually, Fe ions tend to form magnetic coupling between second-nearest neighbors [30]. Thus, the mean domain size is of the nm

scale.

[50] S. Hüfner, S.-H. Yang, B.S. Mun, C.S. Fadley, J. Schäfer, E. Rotenberg, and S.D. Kevan, Phys. Rev. B **61**, 12582 (2000).

[51] G. Levy, R. Sutarto, D. Chevrier, T. Regier, R. Blyth, J. Geck, S. Wurmehl, L. Harnagea, H. Wadati, T. Mizokawa, I.S. Elfimov, A. Damascelli, and G.A. Sawatzky, Phys. Rev. Lett. **109**, 077001 (2012).

[52] D.D. Sell, H.C. Casey, and K.W. Wecht, J. Appl. Phys. **45**, 2650 (1974).

[53] K.W. Edmonds, G. van der Laan, A.A. Freeman, N.R.S. Farley, T.K. Johal, R.P. Campion, C.T. Foxon, B.L. Gallagher, and E. Arenholz, Phys. Rev. Lett. **96**, 117207 (2006).

[54] K.W. Edmonds, N.R.S. Farley, T.K. Johal, G. van der Laan, R.P. Campion, B.L. Gallagher, and C.T. Foxon, Phys. Rev. B **71**, 064418 (2005).

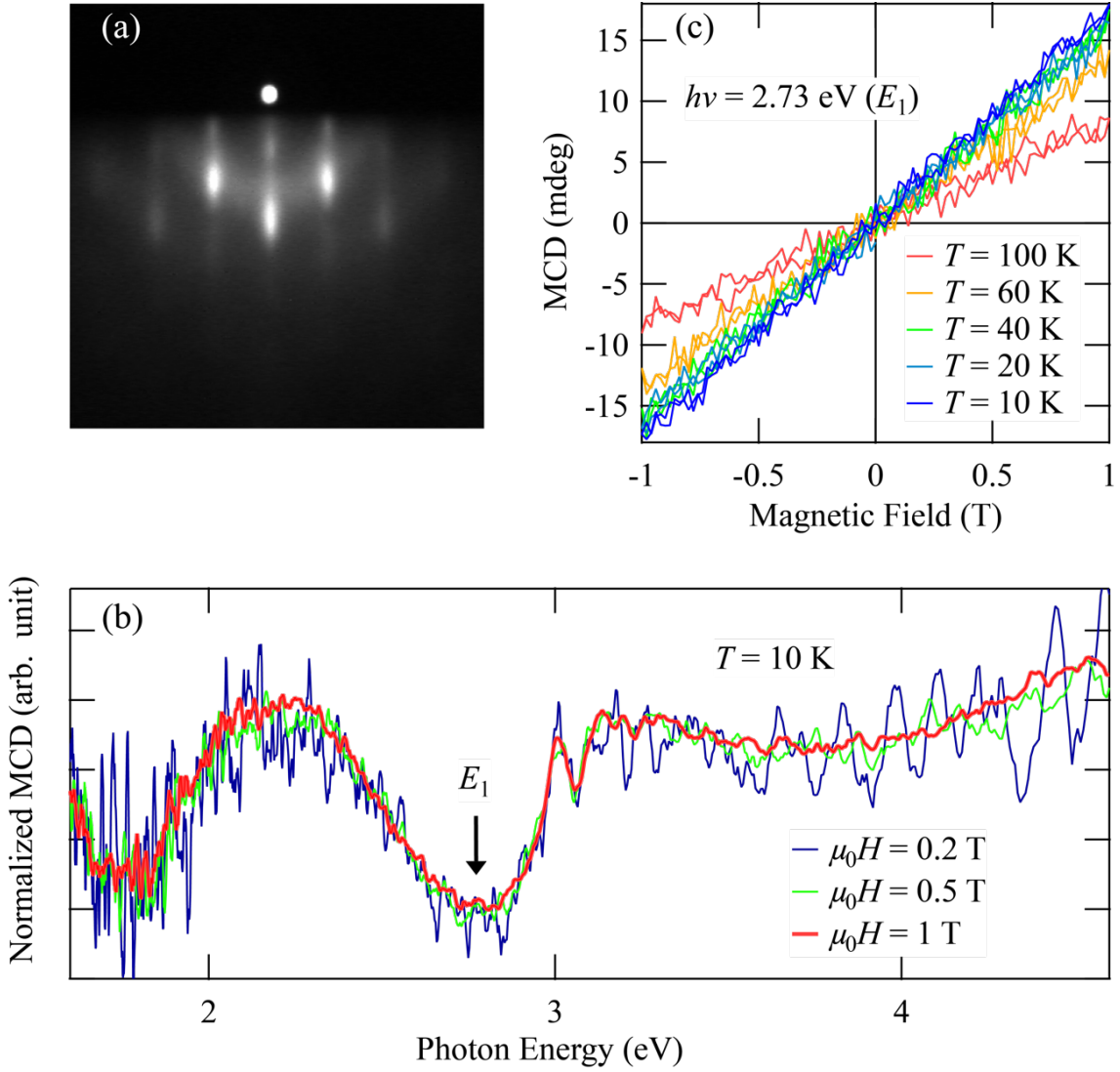

FIG. 1: Characterization of the $(Ga_{0.95}, Fe_{0.05})As$ thin film. (a) RHEED pattern of the $(Ga_{0.95}, Fe_{0.05})As$ film taken along the [-110] azimuth after the MBE growth. (b) Normalized reflection MCD spectra measured at $T = 10$ K under magnetic fields $\mu_0 H = 0.2$ T, 0.5 T, and 1 T applied perpendicular to the film. (c) MCD-$H$ curves measured at $E_1$ ($h\nu = 2.73$ eV).

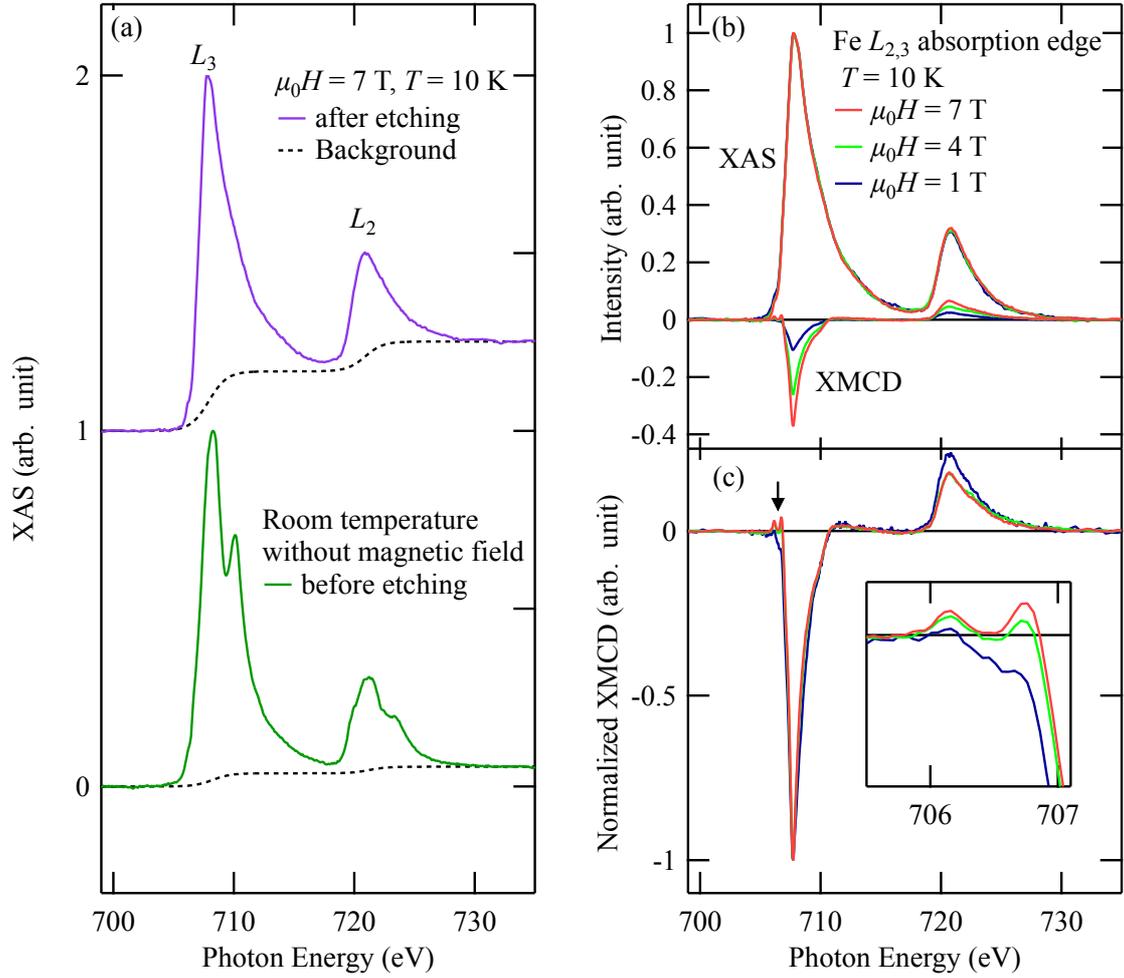

FIG. 2: Fe $L_{2,3}$-edge XAS and XMCD spectra of $(Ga_{0.95},Fe_{0.05})As$ at $T = 10$ K. (a) XAS spectrum of after-etching (purple curve) and before-etching (green curve) samples, where dashed black curves represent the background. (b) and (c) Magnetic-field dependence of XAS and XMCD spectra, respectively. Here, the spectra have been normalized to the height at $h\nu = 708$ eV. In the inset, the XMCD spectra at the $L_3$ edge are enlarged and the pre-edge structures are clearly observed.

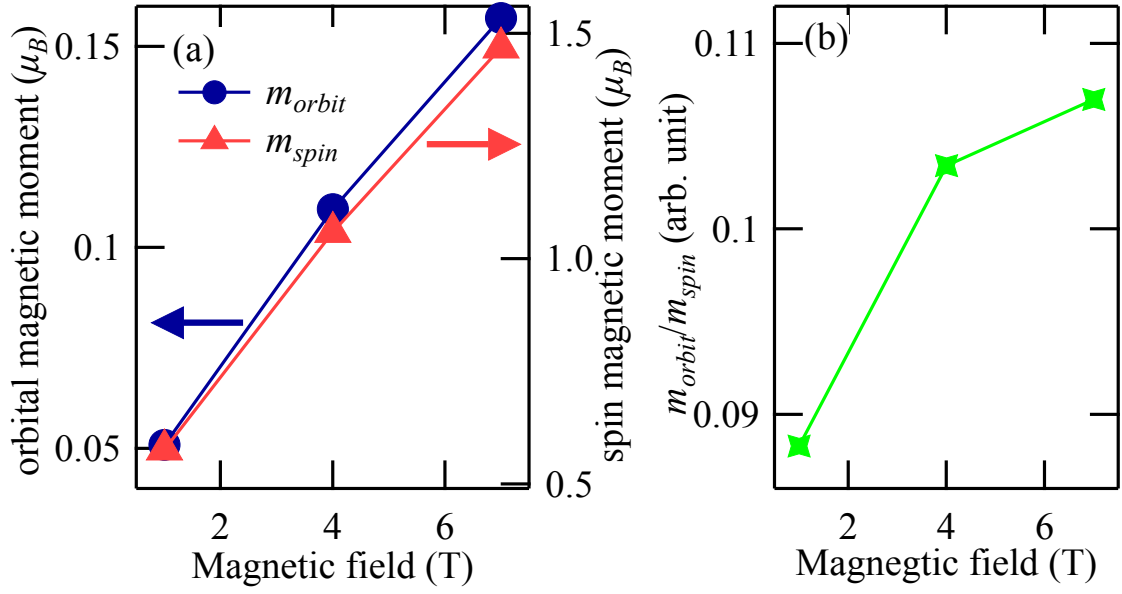

FIG. 3: Spin and orbital magnetic moments estimated using the XMCD sum rules. (a) Magnetic-field dependence of orbital (left axis, round markers) and spin (right axis, triangle markers) magnetic moments. (b) Magnetic-field dependence of the orbital magnetic moment-to- spin magnetic moment ratio.

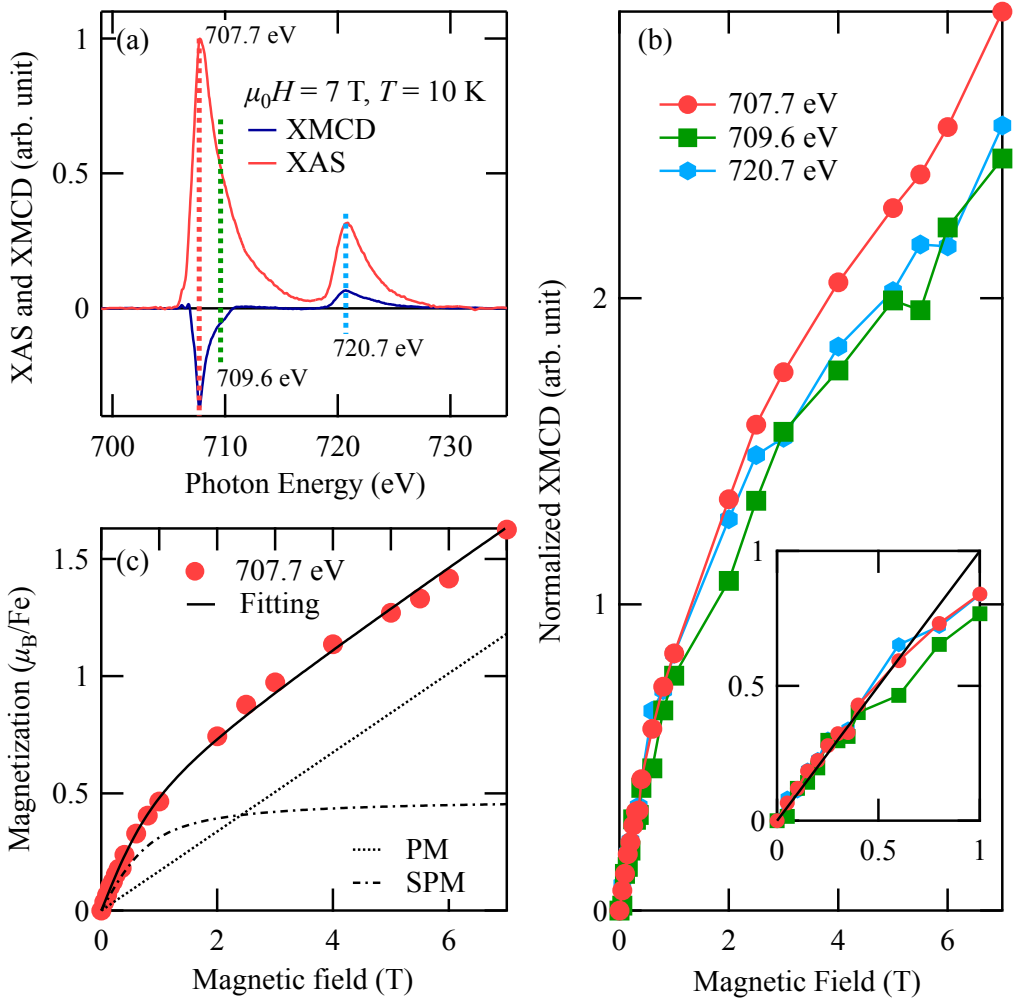

FIG. 4: Magnetic field dependence of the XMCD of $(Ga_{0.95},Fe_{0.05})As$. (a) XAS and XMCD spectra, where the photon energies used for the XMCD-$H$ measurements are indicated by dashed bars. (b) XMCD-$H$ curves taken at various $h\nu$'s, where the XMCD-$H$ curves have been normalized such that the slopes of the curves between 0 T and 0.4 T become the same. In the inset, the normalized XMCD-$H$ curves between 0.1 T and 1 T are enlarged, where the black line represents the XMCD intensity proportional to $\mu_0 H$. (c) Fitting of the magnetization curve taken at $h\nu = 707.7$ eV. The fitting result is shown by the solid black curve. Here, the PM and SPM components of the fitting result are represented by black dotted and dash-dotted curves, respectively.

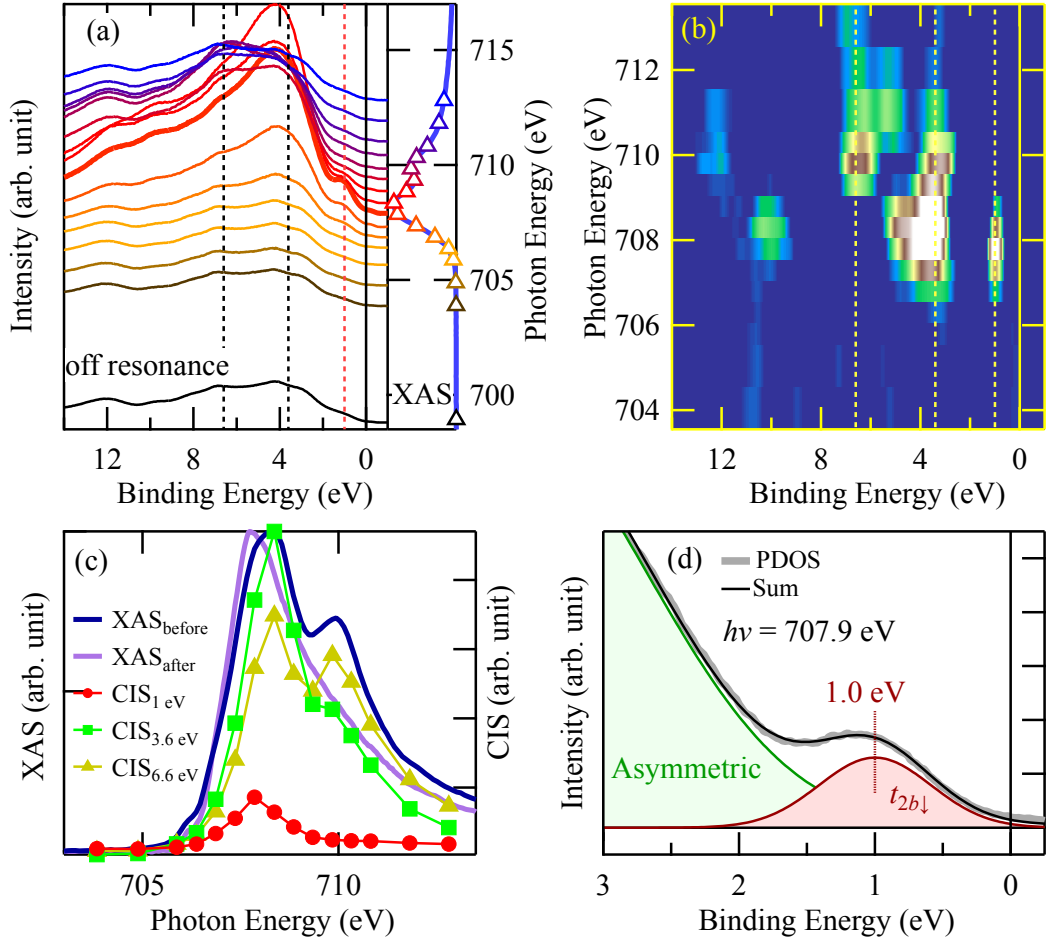

FIG. 5: RPES spectra of a unetched $(Ga_{0.95},Fe_{0.05})As$ thin film. (a) RPES spectra taken at various $hv$'s across the Fe $L_3$ absorption edge. The colors of the spectra correspond to those of the markers on the XAS spectrum in the right panel. The RPES spectrum represented by the thick red curve in the left panel is taken at $hv = 707.9$ eV. (b) Second-derivative image of the RPES spectra, which are obtained by subtracting the off-resonant spectrum (black spectrum in Fig. 2(b)) from the spectra in Fig. 2(b). (c) CIS spectra $E_B = 1.0$ eV, 3.6 eV, and 6.6 eV. For comparison, XAS spectra before and after etching are also plotted. (d) Fitting of the PDOS spectrum taken at $hv = 707.9$ eV. The solid gray, solid black, solid red, solid green curves represent the PDOS spectrum, sum of fittings, an asymmetric Gaussian fitting, and a Gaussian fitting, respectively. Here, the PDOS spectrum is obtained by subtracting the off-resonant spectrum from the RPES spectrum taken at $hv = 707.9$ eV.

## (a) (Ga$_{0.95}$,Fe$_{0.05}$)Sb [Ferromagnetic]

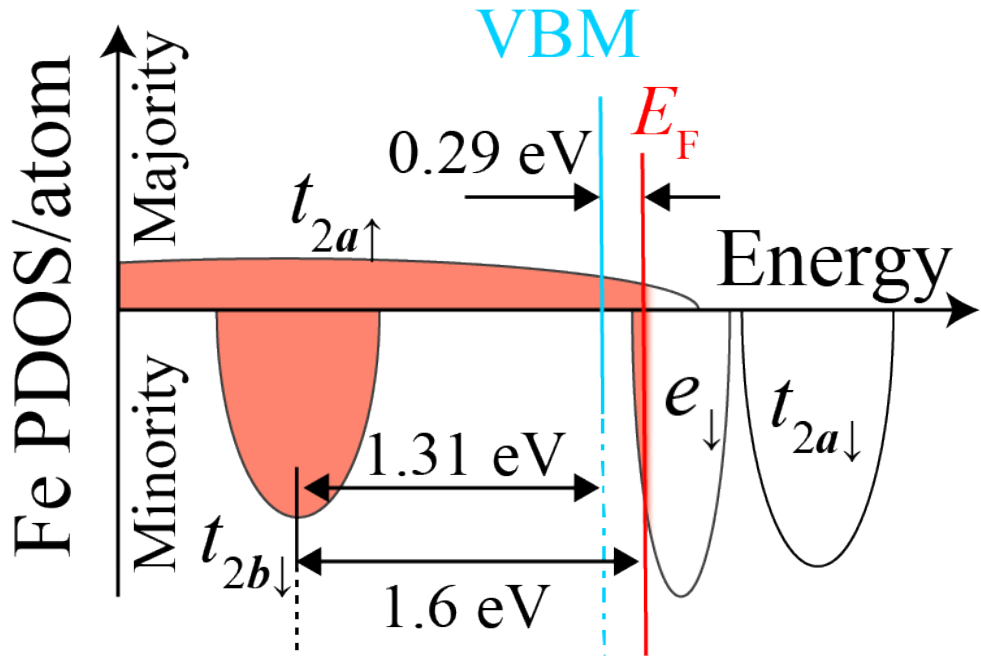

## (b) (Ga$_{0.95}$,Fe$_{0.05}$)As [Paramagnetic]

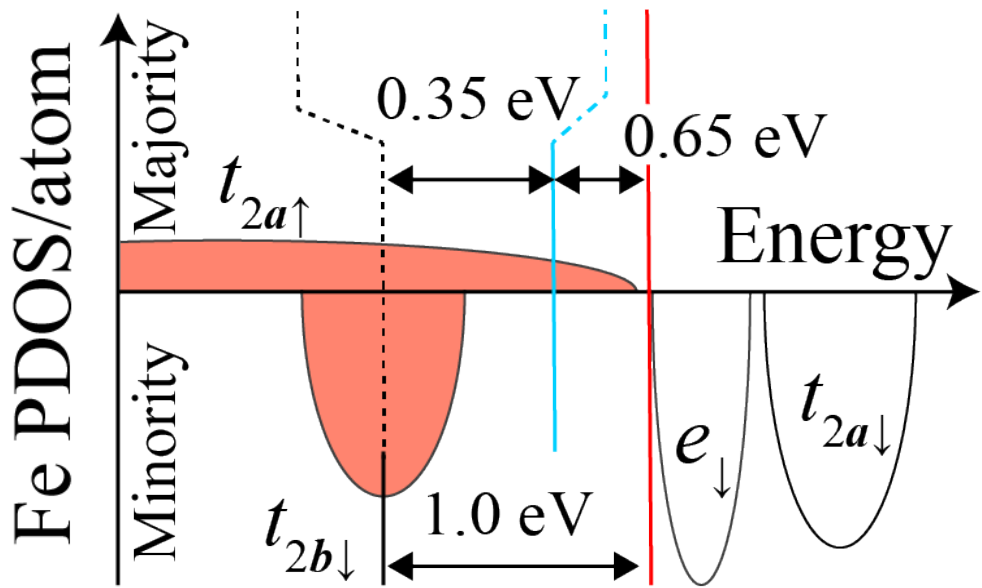

FIG. 6: Schematic energy diagrams of (a) (Ga$_{0.95}$,Fe$_{0.05}$)Sb[27] and (b) (Ga$_{0.95}$,Fe$_{0.05}$).

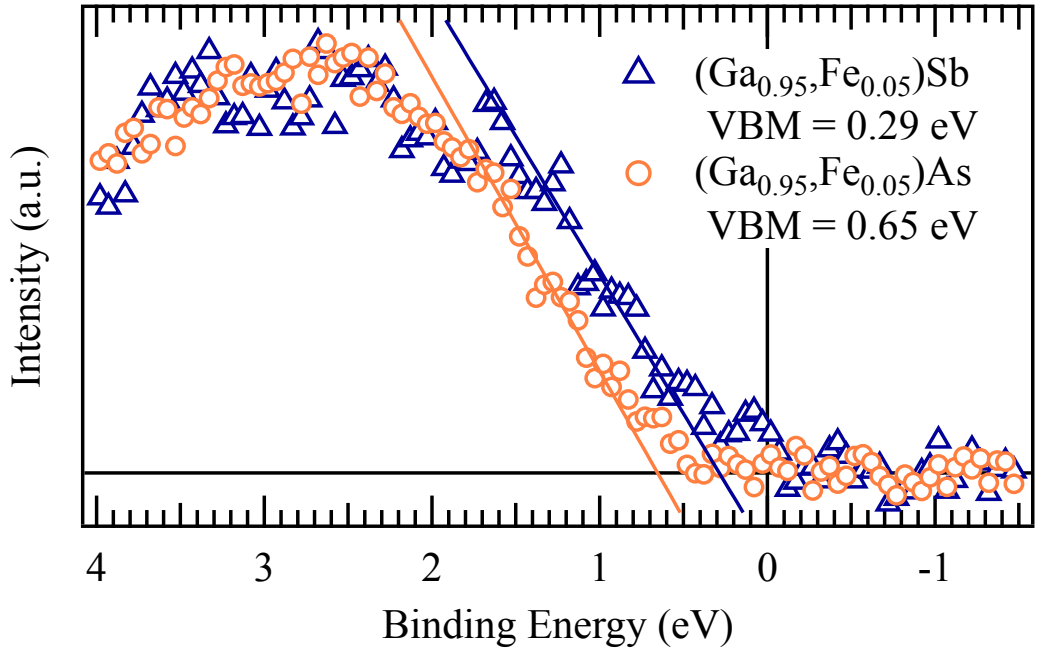

FIG. 7: Mg-$K\alpha$ XPS spectra of $(Ga_{0.95},Fe_{0.05})As$ and $(Ga,Fe)Sb$ near the VBM at room temperature. Here, orange and blue markers represent the XPS spectra of $(Ga_{0.95},Fe_{0.05})As$ and $(Ga,Fe)Sb$, respectively. The solid lines are the results of the linear fitting. The estimated positions of the VBM of $(Ga_{0.95},Fe_{0.05})As$ and $(Ga,Fe)Sb$ are $E_B$ = 0.65 eV and 0.29 eV, respectively.